\documentclass[12pt]{article}

\pdfoutput=1

\usepackage[usenames, dvipsnames]{xcolor}
\usepackage{graphics}
\usepackage{epsfig}
\usepackage{amsfonts}
\usepackage{amssymb}
\usepackage{amsmath}
\usepackage{dsfont}
\usepackage{pifont}
\usepackage{bbm}
\usepackage{multirow}
\usepackage{latexsym}
\usepackage{verbatim}
\usepackage{mcite}
\usepackage{cite}
\usepackage{units}
\usepackage[all]{xy}
\usepackage{cancel}
\usepackage{slashed}
\usepackage{enumerate}
\usepackage{array}
\usepackage{caption}

\definecolor{DarkGreen}{RGB}{18,173,42}
\definecolor{DarkRed}{RGB}{202,0,42}

\DeclareMathAlphabet{\mathbbold}{U}{bbold}{m}{n}

\def\bfone{\relax{\rm 1\kern-.35em 1}}

\newcommand{\be}{\begin{equation}}
\newcommand{\ee}{\end{equation}}
\newcommand{\ben}{\begin{displaymath}}
\newcommand{\een}{\end{displaymath}}
\newcommand{\bea}{\begin{eqnarray}}
\newcommand{\eea}{\end{eqnarray}}

\newcommand{\bean}{\begin{eqnarray*}}
\newcommand{\eean}{\end{eqnarray*}}

\DeclareMathAlphabet{\mathpzc}{OT1}{pzc}{m}{it}

\begin{document}
\pagestyle{plain}


\makeatletter \@addtoreset{equation}{section} \makeatother
\renewcommand{\thesection}{\arabic{section}}
\renewcommand{\theequation}{\thesection.\arabic{equation}}
\renewcommand{\thefootnote}{\arabic{footnote}}


\setcounter{page}{1} \setcounter{footnote}{0}


\begin{titlepage}

\begin{flushright}
UUITP-28/17\\
\end{flushright}

\bigskip

\begin{center}

\vskip 0cm

{\LARGE \bf A swamp of non-SUSY vacua} \\[6mm]

\vskip 0.5cm

{\bf U.~H.~Danielsson, G.~Dibitetto,  \,and\, S.~C.~Vargas}\let\thefootnote\relax\footnote{{\tt \{ulf.danielsson, giuseppe.dibitetto, sergio.vargas\} @physics.uu.se}}\\

\vskip 25pt

{\em Department of Physics and Astronomy, Uppsala University, \\ Box 516, SE-751 20 Uppsala, Sweden \\}

\vskip 0.8cm

\end{center}

\vskip 1cm

\begin{center}

{\bf ABSTRACT}\\[3ex]

\begin{minipage}{13cm}
\small

We consider known examples of non-supersymmetric AdS$_7$ and AdS$_4$ solutions arising from compactifications of massive type IIA supergravity and study their stability, taking into account the coupling between 
closed- and open-string sector excitations. Generically, open strings are found to develop modes with masses below the Breitenlohner-Freedman (BF) bound. 
We comment on the relation with the Weak Gravity Conjecture, and how this analysis may play an important role in examining the validity of non-supersymmetric constructions in string theory.

\end{minipage}

\end{center}

\vfill

\end{titlepage}


\tableofcontents

\section{Introduction}

String theory as a candidate theory of quantum gravity requires UV supersymmetry to be realized for its own consistency. Therefore, the most important challenge that arises in this context is that of
understanding viable dynamical processes that cause supersymmetry breaking at lower energies. This would allow us to incorporate effective descriptions such as cosmic acceleration as well as the 
standard model physics at the electroweak symmetry breaking scale, within a UV-complete theory.

While it was originally argued in \cite{Denef:2004cf} that non-supersymmetric vacua are statistically abundant within the so-called ``string landscape'', the seminal work of \cite{ArkaniHamed:2006dz}
has caused a major paradigm shift in our way of analyzing the intimate relation between UV and IR physics. 
In particular, the weak gravity conjecture (WGC) proposed by the authors of \cite{ArkaniHamed:2006dz} may be viewed as a \emph{bottom-up} criterion that singles out those effective low energy models 
that can be successfully UV-completed in string theory thanks to their consistent spectrum of excitations. Only such models should be referred to as parts of the landscape.

Concretely, the WGC invokes the existence of \emph{microscopic} states in the quantum spectrum of a system, whose mass is smaller than, or equal to, the charge. Recently in \cite{Ooguri:2016pdq}, a stronger 
version of the WGC was proposed according to which the mass/charge inequality may \emph{only} be saturated by BPS objects in a supersymmetric vacuum. The main consequence thereof is that each and 
every non-supersymmetric AdS vacuum must be non-perturbatively unstable w.r.t.~the emission of microscopic charged objects (see also \cite{Freivogel:2016qwc} for a similar analysis).    
Further checks and explict realizations/counterexamples of the above idea in a stringy setup may be found \emph{e.g} in \cite{Cottrell:2016bty,Ooguri:2017njy,Banks:2016xpo}.
  
Still inspired by the idea of studying universal instabilities of non-supersymmetric AdS vacua in string theory, but from a somewhat complementary viewpoint,  a mechanism was
proposed in \cite{Danielsson:2016rmq} that generically predicts tachyons for all non-supersymmetric vacua admitting a so-called \emph{bran e picture} \cite{Kounnas:2007dd}. These tachyons are associated with the extra matter fields
living on the spacetime-filling branes participating in the above brane picture. 
These instabilities crucially rely on the coupling between closed- and open-string modes. Therefore, the above claim should not be viewed to be in conflict with 
the recent analysis in \cite{Danielsson:2016mtx} (and in particular used, \emph{e.g.} in \cite{Apruzzi:2016rny}, for holographic purposes), where some classes of non-supersymmetric AdS extrema in massive
type IIA compactifications are found to be even non-perturbatively stable. This analysis is restricted to the closed-string sector, whereas the main take-home message in \cite{Danielsson:2016rmq}
is that one should worry more about open-string modes when it comes to stability of non-supersymmetric vacua.
The aim of this paper is precisely that of further elaborating this proposal by giving concrete examples where the presence of such tachyons can be explicitly shown. The models chosen here are exactly those
of \cite{Danielsson:2016mtx}, which were argued to be fully stable within the closed-string sector. 

The paper is organized as follows. In section~\ref{sec:AdS7}, we review the $\mathcal{N}=1$, $D=7$ gauged supergravity description of the (non-)supersymmetric $\textrm{AdS}_{7}$ solutions of 
\cite{Blaback:2010sj,Apruzzi:2013yva,Passias:2015gya} found in the context of massive type IIA supergravity. We will start by studying the theory that accounts for all closed-string zero modes and
subsequently move to deriving the D6-probe effective potential directly from the 10D perspective. 
Finally, we will consider the 7D coupling with extra vector multiplets describing the position moduli of spacetime-filling D6-branes and match the resulting mass spectra.  
In section~\ref{sec:AdS4}, we start by reviewing the $\mathcal{N}=4$, $D=4$ gauged supergravity description of the (non-)supersymmetric $\textrm{AdS}_{4}$ solutions of \cite{Behrndt:2004km}
obtained from massive IIA reductions on twisted tori with fluxes. The above supergravity model accounts for all closed-string zero mode excitations. Subsequently, we will propose an extended 4D theory
obtained by the coupling with extra vector multiplets in order to describe position and YM moduli for the spacetime-filling D6-branes.
In both the 7D and the 4D case, we will encounter tachyonic modes within the open-string sector. Possible end-points of these instabilities, and their relevance for a consistent theory of quantum gravity, will be discussed in section~\ref{sec:Discussion}.

\section{Massive type IIA on AdS$_7 \times S^3$}
\label{sec:AdS7}

The goal of this section is to analyze the different behavior of the open-string sector in the supersymmetric and the non-supersymmetric $\textrm{AdS}_{7}$ solutions found in the context of
massive type IIA supergravity on $S^{3}$. To this end, we will first review the form of the 10D solutions and subsequently we will move to their gauged 7D supergravity description.
The closed-string excitations will be accounted for by considering the gravity multiplet coupled to three vector multiplets. 
Then we will move to the study of open-string excitations by calculating the effective potential of spacetime-filling D6 probes. In the non-supersymmetric extremum the position of the
D6-branes will turn out to be tachyonic, while it will be above the BF bound \cite{Breitenlohner:1982jf} in the supersymmetric solution. 
Finally, we will be able to recover exactly the same result by computing the mass matrix of the corresponding 7D gauged supergravity theory when coupled to $N$ extra vector multiplets.

\subsection{Massive type IIA solutions in AdS$_7 \times S^3$}\label{AdS710DsolsSection}
 
The AdS$_7 \times S^3$ solutions of interest can be described in 10D massive IIA supergravity with a background including the RR 1-form $C_{(1)}$, the Romans' mass $F_{(0)}$, the NSNS $B_{(2)}$ field, 
the dilaton $\Phi$ and the metric $g$. We write these as \cite{Passias:2015gya}
\begin{eqnarray}
\mathrm{d}s_{10}^2 &=& \frac{1}{16} \, X^{-1/2} \, e^{2 A} \, \mathrm{d}s^2_{\mathrm{AdS}_7} \, + \, X^{5/2} \, \left[ \mathrm{d}r^2 \,+ \, e^{2 A} \, \frac{ 1-\xi^2}{16w} \left( \mathrm{d}\theta^2 \, + \, \sin^2 \theta \, \mathrm{d}\psi^2 \right) \right] \ , 
\\
\mathrm{d}s^2_{\mathrm{AdS}_7} &=& e^{4z/L} \, \mathrm{d}s^2_{\mathrm{Mkw}_6} \, + \, \mathrm{d}z^2 \ ,
\\
C_{(1)} &=& \frac{1}{4} \, \cos \theta \, e^{A-\Phi_{0}} \, \sqrt{1-\xi^2} \, \mathrm{d} \psi  \ ,
\\
F_{(0)} &=& m \ ,
\\
B_{(2)} &=& - \, \frac{1}{8} \ e^A \, \cos \theta \, \mathrm{d} r \wedge \mathrm{d} \psi \, + \, \frac{e^{2 A} \xi \sqrt{1-\xi^2}  \sin \theta}{32 \left[ \xi^2+X^5 \left(1-\xi^2\right) \right]} \, \mathrm{d} \theta \wedge \mathrm{d} \psi \ ,
\\
\Phi &=& \Phi_{0} \, + \, \frac{1}{2} \, \log \left(\frac{X^{5/2}}{w}\right) \ ,
\end{eqnarray}
where
\begin{eqnarray}
w &\equiv& \xi^2 \,+ \, X^5 \left(1 \, - \, \xi^2 \right) \ ,
\\
L \, &\equiv& \,  8  \sqrt{15} \, X^4 \, (8 X^{10}+8 X^5-1)^{-1/2} \ .
\end{eqnarray}
It is useful to change from the coordinate $r$ to a coordinate $y$ via
\begin{equation}
\mathrm{d}r \, = \, \frac{9}{16}  \, \frac{e^{3 A}}{\sqrt{\beta }} \, \mathrm{d}y \ ,
\end{equation}
which allows one to analytically describe a family of solutions in terms of a function $\beta \, = \, \beta (y)$. In terms of the $y$ coordinate, the north pole of $S^{3}$ is located at $y=-2$.
The case of our interest is the AdS vacuum supported by a stack of D6-branes located at $y=-2$, which corresponds to the following choice
\begin{eqnarray}
\label{sol:D6@y=-2}
\beta &=& \frac{8}{m} \, \left( y-1 \right) \, \left( y+2 \right)^2 \ ,
\\
\xi^2 &=& \, - \, \frac{y \beta'}{ 4 \beta \, - \, y \beta'} \ ,
\\
e^A &=& \, \frac{2}{3} \left( - \frac{\beta'}{y} \right)^{1/4} \ ,
\\
e^{\Phi_{0}} &=& \, \frac{1}{12} \, \left( 4 \beta \, - \, y \beta'  \right)^{-1/2} \, \left( - \frac{\beta'}{y} \right)^{5/4} \ .
\end{eqnarray}
Note that the above background is a complete solution to the set of 10D field equations in appendix~\ref{App:MIIA}, provided that the scalar $X$ satisfies
\be
1\,-\,3X^{5}\,+\,2X^{10} \ = \ 0 \ ,
\ee
which holds for $X\,=\,1$ (SUSY extremum), and $X\,=\,2^{-1/5}$ (non-SUSY extremum).

Since the above solutions are supported by \emph{spacetime-filling} D6-branes, they require the inclusion of a source term on the rhs of one of the BI in \eqref{eq:bianchi},
to yield something of the form of
\be
dF_{(2)} \, - \, F_{(0)} \,\wedge\, H_{(3)} \, = \, N_{(\textrm{D}6)} j_{(3)} \ ,
\ee
where $j_{(3)}$ denotes a 3-form current. Such D6-branes would then fill $\textrm{AdS}_{7}$ and be fully localized at $y=-2$ inside $S^{3}$.

In the next subsection we will review the 7D effective description of the above AdS vacua within $\mathcal{N}=1$ gauged supergravity coupled to three vector multiplets. This will immediately allow us to
identify those with the critical points of the $\textrm{ISO}(3)$-gauged theory found in \cite{Dibitetto:2015bia}.

\subsection{7D gauged supergravity description}
\label{sec:gauged7D_3VM}

Warped compactifications of massive type IIA supergravity on a squashed $S^{3}$ with spacetime-filling O6/D6 sources are known to admit a gauged $\mathcal{N}=1$, $D=7$ supergravity description.
The theory that captures all of the closed-string zero modes is the one obtained through the coupling of the gravity multiplet with three extra vector multiplets.
Such a supergravity model enjoys 
\be
G_{0} \ \equiv \ \mathbb{R}^{+}\,\times\,\textrm{SO}(3,3) 
\notag
\ee
as a global symmetry, and its $64$ bosonic degrees of freedom are arranged into the metric, six vector fields, one two-form potential and ten scalars.
Such 7D degrees of freedom encode the full amount of information concerning the spectrum of zero modes within the orientifold-even sector of closed-string excitations.
The explicit dictionary is given in table~\ref{Table:IIAfields_7D}.
\begin{table}[h!]
\renewcommand{\arraystretch}{1}
\begin{center}
\scalebox{1}[1]{
\begin{tabular}{|c|c|c|}
\hline
IIA fields & $\mathbb{Z}_{2}$-even components & 7D fields  \\
\hline \hline
\multirow{2}{*}{$g_{MN}$} & $g_{\mu\nu}$ & graviton $(\times 1)$\\
\cline{2-3} & $g_{ij}$ & scalars $(\times 6)$ \\
\hline
$B_{MN}$ & $B_{\mu i}$ & vectors $(\times 3)$\\
\hline
$\Phi$ & $\Phi$ & scalar $(\times 1)$\\
\hline
$C_{M}$ & $C_{i}$ & scalars $(\times 3)$\\
\hline
\multirow{2}{*}{$C_{MNP}$} & $C_{\mu\nu\rho}$ & 3-form $(\times 1)$\\
\cline{2-3} & $C_{\mu jk}$ & vectors $(\times 3)$ \\
\hline
\end{tabular}
}
\end{center}
\caption{{\it The effective 7D bosonic field content of type IIA supergravity on a compact 3d manifold. Please note that the orientifold truncation selects the field components which are even w.r.t.
the O6 projection. After dualizing the 3-form into a 2-form, the counting exactly matches the bosonic part of the 7D field content.}} 
\label{Table:IIAfields_7D}
\end{table}

The scalar fields span the following coset geometry
\be
\underbrace{\mathbb{R}^{+}_{X}}_{X}\,\times\,\underbrace{\frac{\textrm{SO}(3,3)}{\textrm{SO}(3)\,\times\,\textrm{SO}(3)}}_{M_{AB}} \ ,
\ee
where $A,\,B,\,\dots$ denote fundamental $\textrm{SO}(3,3)$ indices. The kinetic Lagrangian for the scalar sector reads
\be
\label{Lkin_7D}
\mathcal{L}_{\textrm{kin}}\,=\,-\frac{5}{2}\,X^{-2}\,\left(\partial X\right)^{2} \, + \,\frac{1}{16}\, \partial_{\mu}M_{AB}\, \partial^{\mu}M^{AB} \ .
\ee
The set of consistent embedding tensor deformations that we are interested in comprizes a three-form $f_{ABC}$ parametrizing the gauging of a subgroup of $\textrm{SO}(3,3)$, and a St\"uckelberg-like
mass deformation for the two-form, which we denote by $\theta$. By means of group-theoretical considerations, it is possible to single out the parts of $f$ and $\theta$ which describe all parameters
giving rise to non-trivial contributions to the scalar potential, \emph{i.e.} the $S^{3}$ extrinsic curvature $\Theta_{ij}$, the $H_{(3)}$ flux and the Romans' mass $F_{(0)}$. This results in the embedding tensor/fluxes dictionary spelled out in table~\ref{table:ET/fluxes}.
\begin{table}[h!]
\begin{center}
\begin{tabular}{| c | c |}
\hline
IIA fluxes & $\Theta$ components  \\
\hline \hline
$F_{(0)}$ & $\sqrt{2} \, \tilde{q}$ \\[1mm]
\hline
$H_{ijk}$ & $\frac{1}{\sqrt{2}} \, \theta \, \epsilon_{ijk}$  \\[1mm]
\hline
$\Theta_{ij}$ & $q\,\delta_{ij}$\\[1mm]
\hline
\end{tabular}
\end{center}
\caption{{\it The embedding tensor/fluxes dictionary for the case of massive type IIA reductions on $S^{3}$. The underlying 7D gauging is generically $\textrm{ISO}(3)$, except when 
$q \, =  \, \tilde{q} \, = \, \lambda $, where it degenerates to $\textrm{SO}(3)$ \protect\cite{Dibitetto:2015bia}.} 
\label{table:ET/fluxes}}
\end{table}
With this identification, in
cartesian
coordinates $f_{ABC}$ can be written as
\bea
f_{ABC} \, = \, \frac{1}{4\sqrt{2}} && \hspace{-6mm} \left[  \left( 3 q - \tilde{q} \right) \epsilon_{ABC456} +  \left( 3 q + \tilde{q} \right) \epsilon_{123ABC} + \left(  q - \tilde{q} \right) \left( \epsilon_{1BCA56} + \epsilon_{A2C4B6} -\epsilon_{AB345C} \right)  \right. \nonumber \\
&&
\hspace{-4mm} \left.  +  \left(  q + \tilde{q} \right) \left( \epsilon_{A234BC} + \epsilon_{1B3A5C} -\epsilon_{12CAB6} \right)  \right] \, .
\eea
%
The scalar potential induced by the embedding tensor reads 
\be
\label{V_7D}
\begin{array}{lclc}
V & = & \frac{1}{64}\,\bigg[\frac{X^{2}}{2}\,f_{ABC}f_{DEF}\,\left(\frac{1}{3}\,M^{AD}M^{BE}M^{CF}\,+\,\left(\frac{2}{3}\,\eta^{AD}\,-\,M^{AD}\right)\,\eta^{BE}\eta^{CF}\right) \, + & \\[2mm]
& & + \ \theta^{2}\,X^{-8} \, - \, \frac{2}{3}\sqrt{2}\,X^{-3}\,\theta\,f_{ABC}\,M^{ABC}\bigg] & ,
\end{array}
\ee
where $\eta^{AB}$ denotes the $\textrm{SO}(3,3)$-invariant metric, $M^{AB}$ is the inverse of $M_{AB}$, and
\be
M^{ABC}\,\equiv\,\epsilon^{abc}\,\mathcal{V}_{a}^{\phantom{a}A}\,\mathcal{V}_{b}^{\phantom{a}B}\,\mathcal{V}_{c}^{\phantom{a}C}\ ,
\ee
in terms of the vielbein $\mathcal{V}$ defined through $\mathcal{V}\mathcal{V}^{\textrm{T}}\,=\,M$. Note that the above contraction is only on $\textrm{SO}(3)_{\textrm{timelike}}$.

By performing the specific choice of embedding tensor in table~\ref{table:ET/fluxes} with $q \, =  \, \tilde{q} \, = \, \lambda $, one finds two AdS extrema which exactly coincide with the massive type IIA solutions of the previous subsection. Thanks to their gauged supergravity description, one can now calculate the mass spectrum of the closed-string zero-mode excitations. The results are collected in table~\ref{AdS7ClosedSector}. 
%
\begin{table}[!htbp]
\begin{center}
\begin{tabular}{| c || c | c | c | c | c |}
\hline
\textrm{\textsc{id}} & $X$ & $M_{AB}$ & $m^2$ & SUSY  & $m^2 \overset{?}{\geq} m^{2}_{\textrm{BF}}\,\equiv\,- \frac{3}{5}$  \\
\hline \hline
$1$ & $1$ & $\mathds{1}_{6}$ & \begin{tabular}{@{}c c@{}} $\mathbf{-\frac{8}{15}}$ &  $(\times 1)$ \\ $0$ &  $(\times 3)$ \\ $\frac{16}{15}$ &  $(\times 5)$ \\ $\frac{8}{3}$ &  $(\times 1)$ \end{tabular} & {\color{DarkGreen} \ding{52}} & {\color{DarkGreen} \ding{52}} \\
\hline
$2$ & $2^{-1/5}$ & $\mathds{1}_{6}$ & \begin{tabular}{@{}c c@{}} $0$  & $(\times 8)$ \\ $\mathbf{\frac{4}{5}}$ & $(\times 1)$ \\ $\frac{12}{5}$  & $(\times 1)$ \end{tabular} & {\color{DarkRed} \ding{56}} & {\color{DarkGreen} \ding{52}} \\
\hline
\end{tabular}
\end{center}
\captionof{table}{\it Properties of the two AdS$_7 \times S^3$ solutions in the \emph{closed} sector. The values of the squared masses are normalized to the cosmological constant and the correct 
(non-canonical) kinetic metric is taken into account. The two critical points can be identified by the value of the modulus $X$. The fifth column shows whether the solutions are supersymmetric ({\color{DarkGreen} \ding{52}}) 
or not ({\color{DarkRed} \ding{56}}), while the last one whether \emph{all} the masses of the $1+9$ scalar dof's are above the BF bound ({\color{DarkGreen} \ding{52}}), or not ({\color{DarkRed} \ding{56}}).}
\label{AdS7ClosedSector}
\end{table}

While the above results suggest full perturbative stability for both solutions, regardless of supersymmetry, our next step will be that of following the lines of \cite{Danielsson:2016mtx} and look for perturbative instabilities within the open-string sector. We remind the reader that such a sector cannot be consistently disregarded since D6-branes crucially support the solutions, and hence they naturally carry open strings attached whose dynamics has to be taken into account.  

\subsection{Probe potential}\label{ProbePotSection}

Our scope now is to compute the mass of the scalar modes corresponding to the D6-branes' position. To this end, we will have to evaluate the brane action within the 10D background defined by 
\eqref{sol:D6@y=-2} in the probe approximation, up to quadratic order.
A single D$p$-brane interacts with a given background due to its tension and charge. Action-wise, these two contributions are given by so-called the Dirac-Born-Infeld (DBI) term and the 
 Wess-Zumino (WZ) term, respectively describing the brane's gravitational and electromagnetic interaction with the background fields. 
At first order in $\alpha'$, these can be written in string frame as
\begin{equation}\label{dbiandwz}
S_{\mathrm{D}p} \, = \, - \tau_p \int_{\Sigma} \mathrm{d}^{p+1} \xi \, e^{-\Phi} \sqrt{-\det\left[\left.\left(g+B_{(2)}\right)\right|_{\Sigma}+\mathcal{F}\right]} \, + \, \tau_p \int_{\Sigma} \, \left.C\right|_{\Sigma} \wedge \textrm{e}^{\mathcal{F}} \ ,
\end{equation}
where $\left. ...\right|_{\Sigma}$ denotes the pullback of the corresponding fields over the brane worldvolume $\Sigma$, $\tau_p$ is the tension of the brane, $\mathcal{F}$ is the field strength of the 
worldvolume gauge field and $C \equiv \sum_n C_{(n)}$ is a polyform introducing the formal sum over all RR potentials.

Spacetime-filling D6-branes are localized at the ``north pole'' of the $S^3$ and fluctuations depend only on the motion along the fiber. We have parameterized this direction with the $y$ coordinate, 
which will be now promoted to a field $Y(x^\alpha) = y + 2$ describing the brane excitations around the critical point $y = - 2$. 
The first thing we need to do then, is to expand the 10D fields around $y=-2$. This procedure yields
\be
\begin{array}{lclc}
e^{A} & = & 2\,\left(\frac{2}{3}\right)^{3/4}\,\left(-m\right)^{-1/4}\,\left(y+2\right)^{1/4}\ + \dots & , \\[2mm]
\xi^{2} & = & 1 \, - \, (y+2) \ + \dots & , \\[2mm]
\mathrm{d}r^{2} & = & \left(-54m\right)^{-1/2}\,\left(y+2\right)^{-1/2}\,\mathrm{d}y^{2}\ + \dots & , \\[2mm]
e^{\Phi_{0}} & = & \left(-24m^{3}\right)^{-1/4}\,\left(y+2\right)^{3/4}\ + \dots & , 
\end{array}
\ee
which reveals a D6-singularity at $y=-2$. Secondly, we take all contributions up to second order in $Y$ into account, both in the DBI and WZ action.

For the WZ term, there is only one contribution to $C$ coming from the $C_{(7)}$ that satisfies $\mathrm{d} C_{(7)} = F_{(8)} = \star_{10} \, F_{(2)}$, with $F_{(2)} = \mathrm{d} C_{(1)} + m B_{(2)}$. 
On the other hand, the fundamental contribution to the DBI comes from the warp factor of the AdS$_7$ and the dilaton. 
Expanding Equation (\ref{dbiandwz}) produces then DBI and WZ terms which both have linear contributions in $Y$ that cancel each other \emph{exactly}. Note that this is crucial in order to verify that
$Y=0$ is a critical point of the action.
Up to second order in $Y$, we find the effective Lagrangian to be
\begin{equation}
\mathcal{L}_{\mathrm{D}6} \, = \, - \frac{1}{2} \, K_{\mathrm{D}6} \, \tilde{g}^{\alpha \beta} \, \partial_\alpha Y \, \partial_\beta Y \, - \, V_{\mathrm{D}6} \ ,
\end{equation}
where
\begin{eqnarray}
K_{\mathrm{D}6} &\equiv&  \left.  \tau_6 \, \mathrm{e}^{-\Phi} \, g_{yy} \, \left(\frac{1}{16} X^{-1/2}\right)^{5/2} \, \mathrm{e}^{5A} \right|_{y = -2} \, = \, - \frac{\tau_6}{2 \cdot 3^5 m} \, , \\
V_{\mathrm{D}6} &\equiv& - \frac{\tau_6}{2 \cdot 3^5 m} \, \frac{1}{2 X^3} \left[ \underbrace{2 Y + \left(X^5-1\right) Y^2 }_\textrm{DBI}  \, + \, \underbrace{4 X^5 +2 -2 Y + \left(\frac{1}{2}-X^5\right) Y^2 }_\textrm{WZ} \right]  \\
&=&
 K_{\mathrm{D}6}  \, \frac{1}{2 X^3} \left[ (4 X^5 + 2) \, - \, \frac{1}{2} Y^2 \right] \, ,
\end{eqnarray}	
and $\tilde{g}^{\alpha \beta}$ is the unwarped metric of AdS$_7$. Interestingly, the mass term for the SUSY point comes solely from the WZ term while the non-SUSY point mass term comes from the DBI.

We want to evaluate the mass of the $Y$ mode normalized to the cosmological constant, which is $\Lambda = - 60 L^{-2}$. 
The normalized masses, by also taking the kinetic metric into account, are then given by 
\be
\begin{array}{lclcl}
m^{2}_{Y} & = & \frac{-1}{2 X^3 |\Lambda|} & = & \left\{
\begin{array}{llrl}
-\frac{8}{15} & , & \textrm{for sol.~1} & , \\
-\frac{4}{5} & , & \textrm{for sol.~2} & , 
\end{array}\right.
\end{array}
\ee
where the latter (non-SUSY) solution is \emph{tachyonic}, since the BF bound in 7D is $-\frac{3}{5}$.

\subsection{The coupling to $(3+N)$ extra vector multiplets}

Now we would like to reproduce the same result of the previous subsection by using gauged 7D supergravity as a tool. As we said earlier, the theory with three vector multiplets only contains information 
concerning closed-string excitations. Hence, if we want to capture the dynamics of the D6-brane position moduli, we have to consider an extension of the theory in subsection~\ref{sec:gauged7D_3VM}.
Since each spacetime-filling D6-brane carries a 7D vector multiplet, the correct extended description should be the one where the gravity multiplet is coupled to $(3+N)$ vector multiplets in total, where $N$ is the number of $D6$-branes.

The unique peculariaty of supergravity theories with $16$ supercharges is that of having to coupling to vector multiplets completely determined by supersymmetry requirements, regardless of the amount of
vectors that one wants to include. This means that our new theory will now have 
\be
G_{0} \ \equiv \ \mathbb{R}^{+}\,\times\,\textrm{SO}(3,3+N) 
\notag
\ee
as a global symmetry, and the amount of scalar fields will be 
\be
\underbrace{1 \ + \ 9}_{\textrm{closed string}} \ + \ \underbrace{3N}_{\textrm{open string}} \ ,
\notag
\ee
where the new $3N$ scalars are denoted by $\left\{Y^{Ia}\right\}_{I=1,\,\dots,N}$ and each of them parametrizes the position coordinates $Y^{a}$ of the the $I$-th D6-brane.

As we argued earlier, the form of the kinetic Lagrangian \eqref{Lkin_7D} and of the scalar potential \eqref{V_7D} will be \emph{identical} in this case, except for the fact that now all contractions 
will now be taken over $(6+N)$-dimensional indices, \emph{i.e.} promote
\be
\begin{array}{lclc}
M_{AB} & \longrightarrow & M_{MN} & , \\[2mm]
f_{ABC} & \longrightarrow & f_{MNP} & ,
\end{array}
\ee
where $M\,\equiv\,A\,\oplus\,I$. Now just stick to the embedding tensor deformations of the previous case, \emph{i.e.}
\be
f_{MNP} \ = \ \left\{
\begin{array}{lclc}
f \textrm{ from table~\ref{table:ET/fluxes}} & , & \textrm{ if }\ [MNP]=[ABC] & , \\[2mm]
0 & , & \textrm{ otherwise } & .
\end{array}\right.
\ee

Then we adopt the following explicit parametrization of the $\frac{\textrm{SO}(3,3+N)}{\textrm{SO}(3)\,\times\,\textrm{SO}(3+N)}$ coset \cite{deRoo:2002jf}
\be
\mathcal{V} \, = \, 
\left(\begin{array}{ c  c  c }
e & -e^{-1}\,\left(\mathcal{B}+\frac{1}{2}\mathcal{A}^{\textrm{T}}\mathcal{A}\right) & -e^{-1}\,\mathcal{A}^{\textrm{T}}\\
0 & e^{\textrm{T}} & 0 \\
0 & \mathcal{A} & \mathds{1}_{N}
\end{array}\right) \ ,
\label{param_SO33N}
\ee
where 
\be
e \, \equiv \, 
\left(\begin{array}{ c  c  c }
e^{\phi_{1}/2} & \chi_{1}\,e^{\phi_{2}/2} & \chi_{2}\,e^{\phi_{3}/2}\\
0 & e^{\phi_{2}/2} & \chi_{3}\,e^{\phi_{3}/2} \\
0 & 0 & e^{\phi_{3}/2}
\end{array}\right) \ , \quad \textrm{and } \quad  \mathcal{B} \, \equiv \, 
\left(\begin{array}{ c  c  c }
0 & \chi_{4} & \chi_{5}\\
-\chi_{4} & 0 & \chi_{6}\\
-\chi_{5} & -\chi_{6} & 0
\end{array}\right) 
\ee
complete with $X$ the set of closed-string scalars, while $\mathcal{A} \, \equiv \, \left(Y^{Ia}\right)$ are the open-string scalars. With this parametrization, in the \emph{going to the origin} formalism we chose all the closed-string moduli to evaluate to 0, except for $X$ which becomes 1, for all critical points\footnote{Notice the difference with the description in sections \ref{AdS710DsolsSection} and \ref{ProbePotSection}, in which each critical point corresponds to a different point in the moduli space.}. By plugging the above expressions into the scalar potential through $M\,=\,\mathcal{V}\mathcal{V}^{\textrm{T}}$, we can check that $Y^{Ia}\,=\,0$ is a critical point for all D6 position moduli. In fact, the equations of motion are polynomials in terms of the fluxes and accept an exact algebraic decomposition, allowing us to claim that these are the only critical points with an SO(3) gauging. While the complete scalar potential (\ref{V_7D}) is a relatively complicated expression that depends on the $1+9+3N$ moduli and the fluxes $\lambda$ and $\theta$, we provide here the effective dependence up to quadratic order in the moduli for the $N=1$ case,
\be
V \, = \, \frac{1}{128} \, \left(2 \theta ^2-4 \theta  \lambda -\lambda ^2\right)-\frac{1}{64} \, (X-1) \left(8 \theta ^2-6 \theta  \lambda +\lambda ^2\right) + \frac{1}{2} \, \Pi \cdot \mathcal{M}^2 \cdot \Pi + \, \mathcal{O} \left( \Pi^3 \right) \, , 
\ee
where we have put together the moduli in the vector
\be
\Pi \, = \, \left\{ X-1 , \, \tilde{\phi}_1 , \, \tilde{\phi}_2 , \, \tilde{\phi}_3 , \, \chi_1 , \, \chi_2 , \, \chi_3 , \, \chi_4 , \, \chi_5 , \, \chi_6 , \, Y_1 , \, Y_2 , \, Y_3 \right\} \, ,
\ee
and the non-normalized masses in the matrix
\bea
\mathcal{M}^2 \, = \, \frac{1}{128} \, \mathrm{diag} && \hspace{-6mm} \left[ 2 \left(72 \theta ^2-24 \theta  \lambda -\lambda ^2\right) , \, \lambda  (\lambda -2 \theta ) , \,  \lambda  (\lambda -2 \theta ) , \, \lambda  (\lambda+\theta ) , \right. \\ \nonumber
&& \hspace{-4mm}  2 \lambda  (\lambda - 2 \theta ) , \, 0 , \, 0 , \, 0 , \, 2 \lambda  (\lambda - 2 \theta ) , \, 2 \lambda  (\lambda - 2 \theta ) , \, \, - 2 \theta  \lambda , \\ \nonumber
&& \hspace{-4mm} \left. - 2 \theta  \lambda , \, - 2 \theta  \lambda \right]\, .
\eea
In this presentation, we have made a change of basis for the $\phi_a$, picking instead the mass eigenstates
\be
\tilde{\phi}_1 \, = \, \frac{1}{\sqrt{2}} \left( \phi_1 + \phi_3 \right)   \, ,  \ \ \ \
\tilde{\phi}_2 \, = \, \frac{1}{\sqrt{6}} \left( \phi_1 - 2 \phi_2 - \phi_3 \right)  \, , \ \ \ \
\tilde{\phi}_3 \, = \, \frac{1}{\sqrt{3}} \left(  \phi_1 + \phi_2 - \phi_3 \right) \, .
\ee
As it can be seen, the equations of motion are trivial except for the one for $X$ which is given by the quadratic equation $8 \theta ^2-6 \theta  \lambda +\lambda ^2 = 0$. The only solutions for the complete set of $1+9+3N$ equations of motion are then $\theta = \frac{\lambda}{4}$ and $\theta = \frac{\lambda}{2}$, and correspond to the SUSY and non-SUSY solution, respectively. At the origin, the kinetic lagrangian becomes
\be
\label{Lkin_7D_Origin}
 \mathcal{L}_{\textrm{kin}\,(\textrm{O})} \,=\, \frac{1}{2} \left[ 5 (\partial X)^2 + \frac{1}{4} (\partial \tilde{\phi}_a)^2 + \frac{1}{2} (\partial \chi_m)^2 + \frac{1}{2} (\partial Y_{Ia})^2\right] \, .
\ee
With these, it is straightforward to compute the normalized masses for the moduli: the masses of the closed-string moduli are as shown in table \ref{AdS7ClosedSector} while for the open-string sector we found perfect agreement with the probe calculation, \emph{i.e.}
\be
\begin{array}{lclcl}
m^{2}_{Y_{Ia}} & = & \frac{-4 \theta \lambda}{| 2 \theta ^2-4 \theta  \lambda -\lambda ^2 |} & = & \left\{
\begin{array}{llrl}
-\frac{8}{15} & , & \textrm{for sol.~1} & , \\
-\frac{4}{5} & , & \textrm{for sol.~2} & . 
\end{array}\right.
\end{array}
\ee

\section{Massive type IIA on AdS$_4 \times S^3 \times S^3$}
\label{sec:AdS4}

Let us now move to considering the different physical properties of SUSY and non-SUSY AdS vacua in four dimensions when it comes to open-string degrees of freedom. 
The compactifications of our interest in this context are those of massive type IIA on twisted tori. We will first review the 10D solutions and then give their 4D gauged supergravity description.
Subsequently, motivated by the matching of the gauged supergravity calculation with the 10D one obtained in the previous section, we will extract crucial information concerning the open-string 
dynamics by analyzing the mass spectra of the $\mathcal{N}=4$ theory coupled to extra vector multiplets. The final outcome will be once more the generic appearence of tachyons.\footnote{We have not performed the analysis using the probe potential, but expect the results to agree. It would be interesting to verify this explicitly.}

\subsection{Massive type IIA solutions in AdS$_4 \times S^3 \times S^3$}

The AdS$_4 \times S^3 \times S^3$ solutions of massive type IIA supergravity that we want to discuss here are characterized by an $\textrm{SU}(2)\times\textrm{SU}(2)$ twisted internal geometry, 
NS-NS $H_{(3)}$ flux and R-R $F_{(p)}$ fluxes, with $p\,=\,0,\,2,\,4,\,6$. The twisting is comletely specified by a set of 1-forms $\left\{\eta^{a},\,\eta^{i}\right\}_{a,\,i\,=\,1,\,2,\,3}$ such that
\be
\begin{array}{lclc}
\mathrm{d}\eta^{a} & = & -\frac{1}{2}\,\omega_{bc}^{\phantom{bc}a}\,\eta^{b}\,\wedge\,\eta^{c}\,-\,\omega_{jk}^{\phantom{bc}a}\,\eta^{j}\,\wedge\,\eta^{k} & , \\[2mm]
\mathrm{d}\eta^{i} & = & -\omega_{bk}^{\phantom{bc}i}\,\eta^{b}\,\wedge\,\eta^{k} & ,
\end{array}
\ee
where $\omega_{bc}^{\phantom{bc}a}$, $\omega_{jk}^{\phantom{bc}a}$ and $\omega_{bk}^{\phantom{bc}i}$ are constants.

The explicit 10D background reads \cite{DallAgata:2009wsi}
\begin{eqnarray}
\mathrm{d}s_{10}^2 &=& \mathrm{d}s^2_{\mathrm{AdS}_4} \, + \, \rho^{2} \, \left(\eta^{a}\,\otimes\,\eta^{a} \,+\, \eta^{i}\,\otimes\,\eta^{i} \right) \ , 
\\
\mathrm{d}s^2_{\mathrm{AdS}_4} &=& e^{4z/L} \, \mathrm{d}s^2_{\mathrm{Mkw}_3} \, + \, \mathrm{d}z^2 \ ,
\\
F_{(0)} &=& m \ ,
\\
F_{(2)} &=& f_{2}\,\delta_{ai} \, \eta^{a}\,\wedge\,\eta^{i} \ ,
\\
F_{(4)} &=& f_{4}\,\delta_{ai}\delta_{bj} \, \eta^{a}\,\wedge\,\eta^{b}\,\wedge\,\eta^{i}\,\wedge\,\eta^{j} \ ,
\\
F_{(6)} &=& f_{6}\,\textrm{vol}_{(6)} \ ,
\\
H_{(3)} &=& h_{1}\,\epsilon_{ijk}\,\eta^{i}\,\wedge\,\eta^{j}\,\wedge\,\eta^{k}\,+\,h_{2}\,\epsilon_{abk}\,\eta^{a}\,\wedge\,\eta^{b}\,\wedge\,\eta^{k} \ ,
\\
\Phi &=& \Phi_{0} \ ,
\end{eqnarray}
where $\rho$, $\Phi_{0}$, $m$, $f_{p}$, and $h_{l}$ are all suitable constants that satisfy some algebraic constraints in order for the above background to solve the field equations in appendix~\ref{App:MIIA}.
The above background describes both SUSY and non-SUSY AdS extrema, depending on the choices of the constants in the flux \emph{Ansatz}.
Upon using the dictionary in \cite{Danielsson:2014ria}, these solutions may be interpreted as those in \cite{Aldazabal:2007sn,Koerber:2010rn} obtained via $\textrm{SU}(3)$-structure compactifications.

All solutions of this type, even though consistent with the presence of the following types of spacetime-filling O6-planes, 
\be
\begin{array}{rcccccccccc}
 & 0 & 1 & 2 & 3 & a & b & c & i & j & k \\[2mm]
\hline
\textrm{O}6^{||}: &  \times & \times & \times & \times & \times & \times & \times & - & - & -   \\
\textrm{O}6^{\perp}: & \times & \times & \times & \times & \times & - & - & - & \times & \times 
\end{array}\label{O6_planes}
\ee
have a vanishing O6/D6 tadpole in all directions, with the O6 and D6 cancelling eachother. This implies that they admit an embedding into three different 4D theories, with different amounts of supersymmetry, namely $\mathcal{N}=1$,
 $\mathcal{N}=4$, and $\mathcal{N}=8$. The minimal supergravity description in terms of a flux-induced superpotential \cite{Derendinger:2004jn} is the one valid in the most general case where all O6
contribute to the tadpoles. When only the one corresponding to $\textrm{O}6^{||}$ is non-vanishing, the underlying minimal model may be regarded as a truncation of a gauged 
$\mathcal{N}=4$ supergravity. When there are no contribution from any O6 at all, the effective supergravity description can be obtained as a truncation of a gauged maximal supergravity.
In the next subsection, we will present the effective 4D $\mathcal{N}=4$ description, where the full information regarding the closed-string zero modes consistent with the orientifold involution are 
contained in the theory coupled to six vector multiplets.

\subsection{4D gauged supergravity description}
\label{sec:gauged4D_6VM}

Massive type IIA orientifold reductions on twisted tori with fluxes down to four dimensions are known to admit a 4D half-maximal gauged supergravity description \cite{DallAgata:2009wsi,Dibitetto:2010rg}.
As already mentioned above, one has to consider the coupling of the gravity multiplet with six extra vector multiplets in order to comprize all of the closed-string zero modes.
This supergravity theory enjoys
\be
G_{0} \ \equiv \ \textrm{SL}(2,\mathbb{R})\,\times\,\textrm{SO}(6,6)
\notag
\ee
as a global symmetry, and its complete set of bosonic degrees of freedom includes the metric, $12$ vector fields and $38$ scalars. The dictionary between these 4D degrees of freedom and the 10D 
orientifold-even sector of closed-string excitations can be found in table~\ref{Table:IIAfields_4D}. 
\begin{table}[h!]
\renewcommand{\arraystretch}{1}
\begin{center}
\scalebox{1}[1]{
\begin{tabular}{|c|c|c|}
\hline
IIA fields & $\mathbb{Z}_{2}$-even components & 4D fields  \\
\hline \hline
\multirow{3}{*}{$g_{MN}$} & $g_{\mu\nu}$ & graviton $(\times 1)$\\
\cline{2-3} & $g_{\mu a}$ & vectors $(\times 3)$ \\
\cline{2-3} & $g_{ab}$, $g_{ij}$ & scalars $(\times 12)$ \\
\hline
\multirow{2}{*}{$B_{MN}$} & $B_{\mu i}$ & vectors $(\times 3)$\\
\cline{2-3} & $B_{ai}$ & scalars $(\times 9)$ \\
\hline
$\Phi$ & $\Phi$ & scalar $(\times 1)$\\
\hline
$C_{M}$ & $C_{i}$ & scalars $(\times 3)$\\
\hline
\multirow{4}{*}{$C_{MNP}$} & $C_{\mu ab}$ & vectors $(\times 3)$\\
\cline{2-3} & $C_{\mu ij}$ & vectors $(\times 3)$ \\
\cline{2-3} & $C_{abc}$ & scalar $(\times 1)$ \\
\cline{2-3} & $C_{ajk}$ & scalars $(\times 9)$ \\
\hline
$C_{MNPQR}$ & $C_{abijk}$ & scalars $(\times 3)$\\
\hline
\end{tabular}
}
\end{center}
\caption{{\it The effective 4D bosonic field content of type IIA supergravity on a compact 6d manifold. Please note that the orientifold truncation selects the field components which are even w.r.t.
the O6 projection. The counting exactly matches the on-shell bosonic part of the 4D field content.}} 
\label{Table:IIAfields_4D}
\end{table}

The scalar fields span the following coset geometry
\be
\underbrace{\textrm{SL}(2,\mathbb{R})}_{M_{\alpha\beta}}\,\times\,\underbrace{\frac{\textrm{SO}(6,6)}{\textrm{SO}(6)\,\times\,\textrm{SO}(6)}}_{M_{MN}} \ ,
\ee
where $M,\,N,\,\dots$ denote fundamental $\textrm{SO}(6,6)$ indices. The kinetic Lagrangian for the scalar sector reads
\be
\label{Lkin_4D}
\mathcal{L}_{\textrm{kin}}\,=\,\frac{1}{8}\, \partial_{\mu}M_{\alpha\beta}\, \partial^{\mu}M^{\alpha\beta} \, + \,\frac{1}{16}\, \partial_{\mu}M_{MN}\, \partial^{\mu}M^{MN} \ .
\ee
The set of consistent embedding tensor deformations that we are interested in are all contained within an object in the $(\textbf{2},\textbf{220})$, denoted by $f_{\alpha MNP}$, parametrizing the 
gauging of a subgroup of $\textrm{SO}(6,6)$. By means of group-theoretical considerations, it is possible to single out the parts of $f$ which describe all parameters
giving rise to non-trivial contributions to the scalar potential, \emph{i.e.} the internal curvature, the $H_{(3)}$ flux and the RR-fluxes. This results in the embedding tensor/fluxes dictionary
spelled out in table~\ref{table:ET/fluxes_4D}.
\begin{table}[h!]
\begin{center}
\scalebox{1}[1]{
\begin{tabular}{ | c | c |}
\hline
IIA fluxes & $\Theta$ components \\
\hline
\hline
$F_{aibjck}$ & $ -f_{+ \bar{a}\bar{b}\bar{c}} $ \\
\hline
$F_{aibj}$ & $f_{+ \bar{a}\bar{b}\bar{k}}$  \\
\hline
$F_{ai}$ & $ -f_{+ \bar{a}\bar{j}\bar{k}}$  \\
\hline
$F_{(0)}$ & $f_{+ \bar{i}\bar{j}\bar{k}}$  \\
\hline
\hline
${H}_{ijk} $ & $ -f_{- \bar{a}\bar{b}\bar{c}} $ \\
\hline
${\omega}^{\phantom{bk}c}_{ij}$ & $f_{- \bar{a}\bar{b}\bar{k}}$ \\
\hline
\hline
$ H_{a b k} $ & $f_{+ \bar{a}\bar{b}k}$ \\
\hline
$ \omega^{\phantom{bk}j}_{k a} = \omega^{\phantom{bk}i}_{b k} \,\,\,,\,\,\, \omega_{b c}^{\phantom{bk}a} $ & $f_{+ \bar{a}\bar{j} k}=f_{+ \bar{i}\bar{b} k}\,\,\,,\,\,\,f_{+ a\bar{b}\bar{c}}$ \\
\hline
\end{tabular}
}
\end{center}
\caption{{\it The embedding tensor/fluxes dictionary for the case of massive type IIA reductions on a twisted $\mathbb{T}^{6}$ with gauge fluxes. 
The $\textrm{SO}(6,6)$ index $M$ is further split into its light-cone directions $(a,\,i,\,\bar{a},\,\bar{i})$. Adapted from \protect\cite{Dibitetto:2012ia}.} 
\label{table:ET/fluxes_4D}}
\end{table}

The scalar potential induced by the embedding tensor reads \cite{Schon:2006kz}
\begin{eqnarray}
  V &=& \dfrac{1}{64} \, f_{\alpha MNP} \, f_{\beta QRS} M^{\alpha
  \beta} \left[ \dfrac{1}{3} \, M^{MQ} \, M^{NR} \, M^{PS} + \left(\dfrac{2}{3} \, \eta^{MQ} - M^{MQ} \right) \eta^{NR}
  \eta^{PS} \right]  \notag
 \\
  && - \dfrac{1}{144} \, f_{\alpha MNP} \, f_{\beta QRS} \, \epsilon^{\alpha
  \beta} \, M^{MNPQRS} \, ,
\label{V_4D} 
\end{eqnarray}
where $\eta^{MN}$ denotes the $\textrm{SO}(6,6)$-invariant metric, $M^{MN}$ is the inverse of $M_{MN}$ (same for $M_{\alpha\beta}$), and  $M_{MNPQRS}$ may be again written in terms of the vielbein
 $\mathcal{V}$ defined through $M\,=\,\mathcal{V}\mathcal{V}^{\textrm{T}}$, as
\begin{equation}
M_{MNPQRS}\, \equiv\, \epsilon_{\underline{mnpqrs}}\mathcal{V}_{M}^{\phantom{M}\underline{m}}\mathcal{V}_{N}^{\phantom{M}\underline{n}}\mathcal{V}_{P}^{\phantom{M}\underline{p}}\mathcal{V}_{Q}^{\phantom{M}\underline{q}}\mathcal{V}_{R}^{\phantom{M}\underline{r}}\mathcal{V}_{S}^{\phantom{M}\underline{s}} \ .
\end{equation}
The underlined indices here are time-like rather than light-like, and are related by the change of basis
\begin{equation}  
R = \frac{1}{\sqrt{2}} \left(
\begin{array}{ccc}
 -\mathds{1}_{6} & \mathds{1}_{6} \\
  \phantom{-}\mathds{1}_{6} & \mathds{1}_{6}
\end{array}
\right) \ . 
\end{equation}

By choosing the embedding tensor as in table~\ref{table:ET/fluxes_4D}, one finds a set of (non-)SUSY AdS vacua which exactly coincides with the one described in the previous subsection. 
From the viewpoint of $\mathcal{N}=4$ supergravity, the different sixteen critical points are grouped into four families (1 -- 4), and within each family, four discrete sign choices for the fluxes are 
possible. These choices are labeled by $\left\{s_{1},s_{2}\right\}$, with $s_{i}\,\in\,\mathbb{Z}_{2}$. 
However, these solutions only happen to be all inequivalent within their $\mathcal{N}=1$ description, while within $\mathcal{N}=4$ the $s_{1}$ sign flip is a symmetry. Finally, if viewed as solutions,
of maximal supergravity, both $\mathbb{Z}_{2}$'s are symmetries and one can only talk about four different AdS extrema.

By using half-maximal gauged supergravity as a tool, the spectrum of the closed-string excitations was computed in \cite{Dibitetto:2011gm}, and the result is summarized in 
table~\ref{AdS4ClosedSector}.
\begin{table}[!htbp]
\begin{center}
\begin{tabular}{| c || c | c |}
\hline
\textrm{\textsc{id}} & SUSY ? & $m^2 \overset{?}{\geq} \, m^{2}_{\textrm{BF}} \,\equiv\, - \frac{3}{4}$  \\
\hline \hline 
$1_{(s_1,+)}$ & {\color{DarkGreen} \ding{52}} & {\color{DarkGreen} \ding{52}} \\
\hline
$2_{(s_1,+)}$ & {\color{DarkRed} \ding{56}} & {\color{DarkRed} \ding{56}} \\
\hline
$3_{(s_1,+)}$ & {\color{DarkRed} \ding{56}} & {\color{DarkGreen} \ding{52}} \\
\hline
$4_{(s_1,+)}$ & {\color{DarkRed} \ding{56}} & {\color{DarkGreen} \ding{52}} \\
\hline
\end{tabular}
\ \ \ 
\begin{tabular}{| c || c | c |}
\hline
\textrm{\textsc{id}} &  SUSY ? & $m^2 \overset{?}{\geq} \, m^{2}_{\textrm{BF}} \,\equiv\, - \frac{3}{4}$  \\
\hline \hline 
$1_{(s_1,-)}$ & {\color{DarkRed} \ding{56}} & {\color{DarkGreen} \ding{52}} \\
\hline
$2_{(s_1,-)}$ & {\color{DarkRed} \ding{56}} & {\color{DarkRed} \ding{56}} \\
\hline
$3_{(s_1,-)}$ & {\color{DarkRed} \ding{56}} & {\color{DarkGreen} \ding{52}} \\
\hline
$4_{(s_1,-)}$ & {\color{DarkRed} \ding{56}} & {\color{DarkGreen} \ding{52}} \\
\hline
\end{tabular}
\end{center}
\captionof{table}{\it Properties of the eight different AdS$_4 \times S^3 \times S^3$ solutions within the \emph{closed-string} sector. Inequivalent solutions are labeled according to the sign $s_2$. 
The second column shows whether the solutions are supersymmetric ({\color{DarkGreen} \ding{52}}) or not ({\color{DarkRed} \ding{56}}) and the third whether \emph{all} the masses of the $2+6^2$ dof's are above the 
BF bound ({\color{DarkGreen} \ding{52}}) or if some are below ({\color{DarkRed} \ding{56}}).}
\label{AdS4ClosedSector}
\end{table}

Just as in the previous seven-dimensional case, the analysis within the closed-string spectrum of excitations would let us conclude that there exists non-SUSY, but nevertheless stable, AdS vacua, such
as the ones within the family '3' and '4'. Here, as opposed to the 7D case, one even has the option of interpreting the above vacua as solutions of $\mathcal{N}=8$ supergravity, since they exhibit all 
vanishing tadpoles for spacetime-filling sources. 
This approach was adopted in \cite{Dibitetto:2012ia}, where the full mass spectrum in $\mathcal{N}=8$ was computed for the various solutions. Interestingly, the ones within the family '4' were found to
be completely \emph{tachyon-free}. Even more interestingly,  in \cite{Danielsson:2016rmq}, still within this effective description, they were even found to be \emph{non-perturbatively stable}.

To understand what is going on, one should recall that the only descriptions of the above $\textrm{AdS}_{4}$ solutions that actually makes sense, have just the right amount of D6 
branes parallel to the O6-planes introduced in \eqref{O6_planes}, so as to produce a vanishing tadpole in all directions.
However, given these UV realizations of our solutions, one would have to consider the dynamical degrees of freedom living on all D6's in order to claim stability at the full quantum level.

In the next section, we want to make use of the coupling of half-maximal supergravity with $(6+N)$ extra vector multiplets in order to compute the mass spectrum of the open-string scalar degrees of freedom
living on the $\textrm{D}6^{||}$ sources. Note that these are the only ones that may be consistently incorporated within a theory with $16$ supercharges, while the rest of the open-string modes will 
require further explicit SUSY breaking.

\subsection{The coupling to $(6+N)$ extra vector multiplets}

By reasoning in an analogous way to the 7D case, we expect the $\mathcal{N}=4$ theory coupled to extra $(6+N)$ vector multiplets to capture all information concerning both closed-string dynamics and the 
open-string one, describing the stack of $N$ D6-branes present in the background.
Our new theory contains now $2+6^2+6N$ scalar fields spanning in the coset
\begin{equation}
\frac{\textrm{SL}(2,\mathbb{R})}{\textrm{SO}(2)} \,\times \,\frac{\textrm{SO}(6,6+N)}{\textrm{SO}(6)\times \textrm{SO}(6+N)} \  ,
\end{equation}
where the new $6N$ scalars are divided into $3N$ position moduli describing how each of the D6-branes is placed in the transverse $ijk$ directions, and $3N$ YM moduli accounting for the different ways
of wrapping the YM vector field on each brane along an $S^{1}$ in the $abc$ directions to get an axion.

By adopting a similar parametrization as in \eqref{param_SO33N}, and an analogous prescription for promoting all the previous objects into tensors of 
$\textrm{SL}(2,\mathbb{R})\,\times \,\textrm{SO}(6,6+N)$, one may just use \eqref{V_4D} to compute the effective potential.
Furthermore, by following the analysis of \cite{Borghese:2010ei}, we could evaluate the whole mass spectra of the theory with $(6+N)$ vector multiplets for all the AdS solutions presented earlier here.
The results are shown in tables~\ref{ExtenPlusMasses} and \ref{ExtenMinusMasses}. 
\begin{center}
$
\begin{array}{c c}
\begin{array}[t]{|cc|}
\hline
 \multicolumn{2}{|c|}{1_{(s_{1},+)}} \\
\hline
\hline
 -\frac{2}{3} & (\times 1) \\
 0 & (\times 10) \\
 \frac{1}{3} \left(4-\sqrt{6}\right) & (\times 1) \\
 \frac{1}{18} \left(89-5 \sqrt{145}-\sqrt{606-30 \sqrt{145}}\right) & (\times 5) \\
 \frac{1}{6} \left(13-\sqrt{34}\right) & (\times {\color{blue} 3N}) \\
 \frac{1}{3} \left(4+\sqrt{6}\right) & (\times 1) \\
 \frac{1}{18} \left(89-5 \sqrt{145}+\sqrt{606-30 \sqrt{145}}\right) & (\times 5) \\
 \frac{1}{6} \left(13+\sqrt{34}\right) & (\times {\color{blue} 3N}) \\
 \frac{29}{9} & (\times 3) \\
 \frac{1}{9} \left(47-\sqrt{159}\right) & (\times 1) \\
 \frac{1}{18} \left(89+5 \sqrt{145}-\sqrt{30 \sqrt{145}+606}\right) & (\times 5) \\
 \frac{1}{9} \left(47+\sqrt{159}\right) & (\times 1) \\
 \frac{1}{18} \left(89+5 \sqrt{145}+\sqrt{30 \sqrt{145}+606}\right) & (\times 5) \\
\hline
\end{array} &
\begin{array}[t]{|cc|}
\hline
 \multicolumn{2}{|c|}{2_{(s_{1},+)}} \\
\hline
\hline
 -\frac{4}{5} & (\times 1) \\
 -\frac{2}{5} & (\times 1) \\
 0 & (\times 10) \\
 \frac{1}{15} \left(77-5 \sqrt{145}\right) & (\times 5) \\
 \frac{8}{5} & (\times {\color{blue} 3N}) \\
 2 & (\times 1) \\
 \frac{2}{15}\left(31-\sqrt{145}\right) & (\times 5) \\
 \frac{46}{15} & (\times 3) \\
 \frac{16}{5} & (\times {\color{blue} 3N}) \\
 \frac{64}{15} & (\times 1) \\
 \frac{2}{15} \left(31+\sqrt{145}\right) & (\times 5) \\
 \frac{20}{3} & (\times 1) \\
 \frac{1}{15} \left(77+5 \sqrt{145}\right) & (\times 5) \\
\hline
\end{array} \\
\begin{array}[t]{|cc|}
\hline
 \multicolumn{2}{|c|}{3_{(s_{1},+)}} \\
\hline
\hline
 0 & (\times 11) \\
 2 & (\times 2) \\
 \frac{1}{3} \left(19-\sqrt{145}\right) & (\times 10) \\
 3 & (\times {\color{blue} 6N}) \\
 \frac{14}{3} & (\times 3) \\
 \frac{20}{3} & (\times 2) \\
 \frac{1}{3} \left(19+\sqrt{145}\right) & (\times 10) \\
\hline
\end{array} &
\begin{array}[t]{|cc|}
\hline
 \multicolumn{2}{|c|}{4_{(s_{1},+)}} \\
\hline
\hline
 0 & (\times (16+{\color{blue} 3N})) \\
 \frac{4}{3} & (\times 6) \\
 2 &(\times  4) \\
 \frac{2}{3} \left(\sqrt{5}+1\right) & (\times {\color{blue} 3N}) \\
 \frac{8}{3} & (\times 5) \\
 6 & (\times 6) \\
 \frac{20}{3} & (\times 1) \\
\hline
\end{array} \\
\end{array} 
$
\captionof{table}{\it Normalized masses for the solutions in the extended theory with $s_2  =  +$, in increasing order. The masses related to the open-string sector are shown in blue.}
\label{ExtenPlusMasses}
\end{center}

\begin{center}
$
\begin{array}{c c}
\begin{array}[t]{|cc|}
\hline
 \multicolumn{2}{|c|}{1_{(s_{1},-)}} \\
\hline
\hline
-\frac{5}{6} & (\times {\color{blue} 3N}) \\
 -\frac{2}{3} & (\times 1) \\
 0 & (\times 10) \\
 \frac{1}{3} \left(4-\sqrt{6}\right) & (\times 1) \\
 \frac{1}{18} \left(89-5 \sqrt{145}-\sqrt{606-30 \sqrt{145}}\right) & (\times 5) \\
 \frac{1}{3} \left(4+\sqrt{6}\right) & (\times 1) \\
 \frac{1}{18} \left(89-5 \sqrt{145}+\sqrt{606-30 \sqrt{145}}\right) & (\times 5) \\
 \frac{5}{2} & (\times {\color{blue} 3N}) \\
 \frac{29}{9} & (\times 3) \\
 \frac{1}{9} \left(47-\sqrt{159}\right) & (\times 1) \\
 \frac{1}{18} \left(89+5 \sqrt{145}-\sqrt{30 \sqrt{145}+606}\right) & (\times 5) \\
 \frac{1}{9} \left(47+\sqrt{159}\right) & (\times 1) \\
 \frac{1}{18} \left(89+5 \sqrt{145}+\sqrt{30 \sqrt{145}+606}\right) & (\times 5) \\
\hline
\end{array} &
\begin{array}[t]{|cc|}
\hline
 \multicolumn{2}{|c|}{2_{(s_{1},-)}} \\
\hline
\hline
 -\frac{4}{5} & (\times (1+{\color{blue} 3N})) \\
 -\frac{2}{5} & (\times 1) \\
 0 & (\times 10) \\
 \frac{1}{15} \left(77-5 \sqrt{145}\right) & (\times 5) \\
 \frac{8}{5} & (\times {\color{blue} 3N}) \\
 2 & (\times 1) \\
 \frac{2}{15} \left(31-\sqrt{145}\right) & (\times 5) \\
 \frac{46}{15} & (\times 3) \\
 \frac{64}{15} & (\times 1) \\
 \frac{2}{15} \left(31+\sqrt{145}\right) & (\times 5) \\
 \frac{20}{3} & (\times 1) \\
 \frac{1}{15} \left(77+5 \sqrt{145}\right) & (\times 5) \\
\hline
\end{array} \\
\begin{array}[t]{|cc|}
\hline
 \multicolumn{2}{|c|}{3_{(s_{1},-)}} \\
\hline
\hline
 -1 & (\times {\color{blue} 3N}) \\
 0 & (\times 11) \\
 2 & (\times 2) \\
 \frac{1}{3} \left(19-\sqrt{145}\right) & (\times 10) \\
 3 & (\times {\color{blue} 3N}) \\
 \frac{14}{3} & (\times 3) \\
 \frac{20}{3} & (\times 2) \\
 \frac{1}{3} \left(19+\sqrt{145}\right) & (\times 10) \\
\hline
\end{array} &
\begin{array}[t]{|cc|}
\hline
 \multicolumn{2}{|c|}{4_{(s_{1},-)}} \\
\hline
\hline
 \frac{2}{3} \left(1-\sqrt{5}\right) & (\times {\color{blue} 3N}) \\
 0 & (\times (16+{\color{blue} 3N})) \\
 \frac{4}{3} & (\times 6) \\
 2 & (\times 4) \\
 \frac{8}{3} & (\times 5) \\
 6 & (\times 6) \\
 \frac{20}{3} & (\times 1) \\
\hline
\end{array} \\
\end{array} 
$
\captionof{table}{\it Normalized masses for the solutions in the extended theory with $s_2  =  -$, in increasing order. The masses related to the open-string sector are shown in blue.}
\label{ExtenMinusMasses}
\end{center}
As one can see from the mass spectra given in the tables, the ``wrong'' sign choice $s_{2}=-$, which turns sol.~1 into a non-supersymmetric one, introduces tachyonic modes within the open-string sector
living on the $\textrm{D}6^{||}$. Moreover, it also turns out to produce similar instabilities in all four families. All of these solutions should therefore be removed from the ``landscape'' of
non-SUSY AdS vacua. However, no analogous conclusion can be drawn for the non-SUSY critical points corresponding to the $s_{2}=+$ sign choice, where no tachyons show up in the open-string sector living
on the $\textrm{D}6^{||}$. On the other hand, one would have to consistently include the rest of the open-string modes associated with $\textrm{D}6^{\perp}$, before claiming full stability. 
This should be done within the framework of $\mathcal{N}=1$ supergravity in four dimensions, and hence it requires some extra work to fix the form of the coupling between this new sector and the rest of the theory. 
Our expectations are that new tachyons will appear there for this other sign choice rather than in the sector considered in our paper, such in a way that it would be impossible to get rid of both tachyonic
sectors at the same time once you give up on SUSY.

\section{Discussion}
\label{sec:Discussion}
Assessing the (non-)existence of a non-supersymmetric AdS landscape in string theory is an issue of utmost importance, both in order to test it as a consistent theory of quantum gravity, and in order to
discuss holography in a more general context where one no longer relies on supersymmetry.
With the above motivations in mind, two different classes of massive type IIA compactifications were studied in this paper, which are known to give rise to non-supersymmetric AdS extrema. 
Both of the above examples were already known in the literature and admit a lower-dimensional gauged supergravity effective description capturing the closed-string zero-mode excitations thereof.
These non-supersymmetric vacua were found to be perturbatively \cite{Dibitetto:2012ia} as well as non-perturbatively \cite{Danielsson:2016rmq} stable within the closed-string sector by 
exploiting such a gauged supergravity description.

However, according to the proposal in \cite{Danielsson:2016mtx}, non-supersymmetric AdS solutions generically develop perturbative instabilities associated with the presence of extra matter living on 
spacetime-filling objects. This idea generalizes the typical instabilites discussed in \cite{Aretakis:2011ha,Aretakis:2011hc,Lucietti:2012xr}, for the $\textrm{AdS}_{2}\times S^{2}$ near-horizon geometry of a Reissner-Nordstr\"om black hole due to the presence of extra
fields fluctuating around the horizon. In particular, according to the brane picture, the  $\textrm{AdS}_{7}$ solutions we have studied can be thought of as part of  $\textrm{AdS}_{7}\times S^{2}$ near-horizon geometries of black five dimensional objects in the world volume of $D8$-branes.

In our present work we took steps towards verifying this picture by explicitly adding the open string degrees of freedom to the theory. We considered the coupling between the closed-string sector of excitations around the non-supersymmetric AdS vacua and the
open-string sector living on the spacetime-filling branes supporting the vacuum. We computed the mass spectra of the open-string excitations and found indications for universal tachyonic modes. In the case of the $\textrm{AdS}_{7}$ solutions, the effective description is given by $\mathcal{N}=1$, $D=7$ gauged supergravity coupled to three vector multiplets. The solutions are all supported by a 
stack of $N$ spacetime-filling D6-branes. The calculation of the masses of the $3N$ position moduli was carried out in two independent ways. The first one consists in treating D6-branes as probes within the
$\textrm{AdS}_{7}\times S^{3}$ background and computing the effective probe potential for the position moduli from the DBI and WZ brane actions. 
The second approach is that of considering the coupling of the above 7D gauged supergravity theory with $N$ extra vector multiplets and identify the correct masses for the new modes from the universal form 
of the scalar potential for theories with sixteen supercharges. The two methods give full agreement on the final answer, \emph{i.e.} the open-string modes are tachyonic in the non-supersymmetric AdS extremum,
which should be then no longer be considered as part of a non-SUSY ``string landscape''. 

Moving to the $\textrm{AdS}_{4}$ case, the different critical points may be embedded within various effective 4D supergravities. This is due to a vanishing tadpole for spacetime-filling O6/D6 sources in
four different directions. Regarded as backgrounds with no orientifolds, and hence no branes, they admit an embedding in $\mathcal{N}=8$ supergravity, and as such, they are fully stable. 
However, thinking along the lines of \cite{Green:2007zzb}, one could argue that the decoupling limit between closed and open strings, which would yield $\mathcal{N}=8$ supergravity, actually is 
\emph{inconsistent} at the level of quantum gravity. The alternative, and correct, description for all compactifications yielding theories with sixteen supercharges, should then include
parallel O-planes and D-branes with vanishing net charge. This agrees with what one expects from anomaly cancelation requirements for type I supergravity in ten dimensions, where its brane realization is
given by O9-planes and D9-branes. By applying T-dualities, this can be related to all backgrounds with O$p$/D$p$ sources. 
Our conclusion is that these open-string instabilities are once again universally present whenever giving up on supersymmetry. 

The first natural issue to be addressed concerns end-point of such an instability. Since in both of the cases analyzed here, the tachyons are related to D6-brane physics, we expect this to be realized
through brane polarization via the Myers' effect \cite{Myers:1999ps}. More concretely, any other positions of the D6-branes in internal space than their original one, involves a wrapping of a 2-cycle of
finite size where they have polarized into D8-branes \cite{Danielsson:2016cit}. This is in line with the supporting $D8$-brane developing a the near horizon $\textrm{AdS}_{7}\times S^{2}$ throat. We checked explicitly that there is no other real minimum of the effective potential when the closed-string moduli are held fixed.
The most natural option is that there could be a new solution crucially involving non-zero $\mathcal{F}$ gauge flux representing dissolved D6-brane charge, in the spirit of the non-supersymmetric 
$\textrm{AdS}_{7}$ solutions found in \cite{Junghans:2014wda}. On the other hand, these constructions heavily rely on the interplay between closed and open strings and hence they are genuinely stringy, since
such competition may only occur at finite $\alpha'$. This makes the actual stability of those solutions an extremely delicate issue that we wish to come back to in the future. In fact, the application of the WGC advocated in  \cite{Danielsson:2016mtx} suggest that they indeed are unstable.

Finally, the other relevant issue that arises from our present analysis, concerns the impact of our results on non-supersymmetric holographic constructions. 
In \cite{Ooguri:2016pdq}, the instabilities were viewed as a direct consequence of the WGC, and argued to destroy the holographic dual by shrinking the lifetime of the boudary CFT to zero.
Our analysis suggests that the issue is much more subtle since the instabilities are only visible when retaining the coupling between open and closed strings.
As suggested in \cite{Danielsson:2016mtx}, the holographic correspondence might then work in a limit where these two sectors can be consistently decoupled from each other. This is not possible in string theory, which could imply that non-supersymmetric duality never is exact but breaks 
down at small scales. We therefore believe that the fate of non-supersymmetric holography needs to be investigated further. 

\section*{Acknowledgments}

We would like to thank Thomas Van Riet and Suvendu Giri for very stimulating discussions. The work of the authors is supported by the Swedish Research Council (VR).


\appendix

\section{Massive type IIA supergravity} \label{App:MIIA}
\noindent In this appendix we review our working conventions concerning \textit{massive} type IIA supergravity in ten dimensions. The bosonic part of its string-frame Lagrangian is given by
\be
\label{action_IIA}
S_{\textrm{IIA}} \ = \ \frac{1}{2\kappa_{10}^{2}} \, \displaystyle\int d^{10}x \, \sqrt{-g} \, \left[e^{-2\Phi}\Big(R \, + \, 4  (\partial\Phi)^{2} 
\, - \, \frac{1}{2} |H_{(3)}|^{2} \Big) \, - \, \frac{1}{2} \sum\limits_{p=0,2,4} |F_{(p)}|^{2}\right] \, + \, S_{\textrm{top.}} \ , 
\ee
where $\,|H_{(3)}|^{2} \ \equiv \ \frac{1}{3!} \, H_{(3)MNP}{H_{(3)}}^{MNP}\,$ and $\,|F_{(p)}|^{2} \ \equiv \ \frac{1}{p!} \, F_{(p)M_{1}\dots M_{p}}{F_{(p)}}^{M_{1}\dots M_{p}}$ with $M=0,...,9$. The above action contains a topological term of the form
\be
\begin{array}{lcl}
S_{\textrm{top.}} & = & -\frac{1}{2} \, \displaystyle\int \left(B_{(2)} \wedge dC_{(3)} \wedge dC_{(3)} \, - \, 
\tfrac{1}{3} F_{(0)} \wedge B_{(2)} \wedge B_{(2)} \wedge B_{(2)} \wedge dC_{(3)} \right. \\[2mm]
& + & \left.\frac{1}{20} F_{(0)} \wedge F_{(0)} \wedge B_{(2)} \wedge B_{(2)} \wedge B_{(2)} \wedge B_{(2)} \wedge B_{(2)} \right) \ .
\end{array}
\ee

\noindent We will now review here set of 10d equations of motion (EOM) and Bianchi identities (BI) which follow from the action \eqref{action_IIA}. 
The equations of motion for $B_{(2)}$, $C_{(1)}$ and $C_{(3)}$ are respectively given by
\be \label{eq:fluxeom}
\begin{array}{rclcccc}
d \left(e^{-2 \Phi }\ast_{10} H_{(3)}\right) & = & 0 &  & & & , \\[2mm]
\left(d \,+\, H_{(3)}\wedge \right)(\ast_{10} F_{(p)}) & = & 0 &  & & (p\,=\,2,\,4) & , \\[2mm]
\end{array}
\ee
whereas the one for the 10D dilaton $\Phi$ reads
\be
\label{eq:dileom}
\begin{array}{rrclc}
\square \Phi \, - \, \left|\partial \Phi \right|^2 \, + \, \frac{1}{4} R \, - \, \frac{1}{8}\left|H_{(3)}\right|^2 & = & 0 & ,
\end{array}
\ee
where $R$ is the 10D scalar curvature and $\square$ is the ten-dimensional (curved) Laplacian operator. The 10D Einstein equations take the standard form\footnote{Note that we have set $\kappa_{10}\,=\,1$.}
\be
\label{10s_Einstein}
\begin{array}{rrclc}
R_{MN} \ - \ \frac{1}{2} \, T_{MN} & = & 0 & ,
\end{array}
\ee
where the symmetric energy-momentum tensor $T_{MN}$ is defined as
\begin{eqnarray}
T_{MN} & = & e^{2\Phi} \,\sum\limits_{p}  \left(  \frac{p}{p!} \,  F_{(p) M M_{1}\dots M_{p-1}}  F_{(p)N}^{\phantom{(p)N}M_{1}\dots M_{p-1}} \, - \, \frac{p-1}{8} \,  g_{MN} \,  |F_{(p)}|^{2}\right)  + \\[2mm]
& + & \Big( \frac{1}{2} \, H_{(3)M PQ} H_{(3)N}^{\phantom{(3)M}PQ} - \frac{1}{4} \,  g_{MN} \,  |H_{(3)}|^{2} \Big) \ - \ \Big(4 \nabla_{M} \nabla_{N} \Phi + \frac{1}{2} \, g_{MN} \left(\square \Phi -2\left|\partial \Phi |^2\right.\right)\Big)  \ ,\notag
\end{eqnarray}
where $\nabla_{M}$ is the covariant derivative w.r.t. the Levi-Civita connection. The trace part of the Einstein equation,
\begin{equation}
R\,-\,5|\partial \Phi |^2\,+\,\frac{9}{2} \,\square \Phi \,-\,\frac{1}{4}|H_{(3)}|^2\,-\,\frac{1}{8} \left(e^{2 \Phi }\right) \sum _p (5-p)|F_{(p)}|^2 \, = \, 0 \ ,
\end{equation}
The (modified) BI instead are given by
\be \label{eq:bianchi}
\begin{array}{rclc}
d H_{(3)} & = & 0 & , \\[2mm]
dF_{(0)} & = & 0 & , \\[2mm]
dF_{(2)} \ - \ F_{(0)} \,\wedge\, H_{(3)} & = & 0 & , \\[2mm]
dF_{(4)} \ + \ F_{(2)} \, \wedge \, H_{(3)} & = & 0 & .
\end{array}
\ee

\bibliography{Swampland}
\bibliographystyle{utphys}

\end{document}